\documentclass[prc,floatfix,groupedaddress,nofootinbib,showpacs,preprintnumbers,
amsmath,amssymb,amsfonts,superscriptaddress,widetable] {revtex4-1}
\usepackage{graphicx}
\usepackage{dcolumn}
\usepackage{mathrsfs}
\usepackage{bm}
\usepackage{float}
\usepackage{soul}
\usepackage[usenames]{color}

\newcommand{\rhoz}{\rho_{\raisebox{-0.75pt}{\tiny 0}}}
\newcommand{\epsz}{\varepsilon_{\raisebox{-0.75pt}{\tiny 0}}}

\newcommand{\gammai}[1]{\gamma^{\raisebox{-1.0pt}{\scriptsize{#1}}}}

\begin{document}

\title{Nuclear breathing mode in neutron-rich Nickel isotopes:\\
sensitivity to the symmetry energy and the role of the continuum}
\author{J. Piekarewicz}
\email{jpiekarewicz@fsu.edu}
\affiliation{Department of Physics, Florida State University, Tallahassee, FL 32306}
\date{\today}
\begin{abstract}
\begin{description}
\item[Background] In this new era of radioactive beam facilities, the 
discovery of novel modes of excitation in nuclei far away from stability
represents an area of intense research activity. In addition, these 
modes of excitation appear to be sensitive to the uncertain
density dependence of the symmetry energy. 
\item[Purpose] To study the emergence, evolution, and nature of 
both the soft and giant isoscalar monopole modes as a function 
of neutron excess in three unstable Nickel isotopes: 
${}^{56}$Ni, ${}^{68}$Ni, and ${}^{78}$Ni.  
\item[Methods] The distribution of isoscalar monopole strength is
computed in a relativistic random-phase approximation using 
several accurately calibrated effective interactions. In particular, 
a non-spectral Green's function approach is adopted that allows 
for an exact treatment of the continuum without any reliance on 
discretization. The discretization of the continuum is neither 
required nor admitted. 
\item[Results] In the case of ${}^{56}$Ni, the lack of low-energy
strength results in a direct correlation between the centroid energy
of the giant monopole resonance and the incompressibility coefficient 
of symmetric nuclear matter. In contrast, the large neutron excess 
in both ${}^{68}$Ni and ${}^{78}$Ni generates a significant, yet relatively 
featureless, amount of low-energy strength that is driven by transitions 
into the continuum. Moreover, the evolution of monopole strength with
neutron excess displays sensitivity to the density dependence of 
the symmetry energy.  
\item[Conclusions] Our results suggest that future measurements of the
distribution of isoscalar monopole strength at radioactive beam facilities 
using a very long chain of both stable and unstable isotopes could place 
important constraints on the equation of state of neutron-rich matter and 
ultimately on the properties of neutron stars. However, given the nature 
of the low-energy monopole excitations, a proper treatment of the 
continuum is essential.
\end{description}
\end{abstract}

\maketitle

\section{Introduction}
\label{intro}
Fundamental new discoveries at radioactive beam facilities all over
the world have led to a paradigm shift in nuclear structure. Core
concepts that have endured the test of time, such as the traditional
magic numbers, are being revisited and revised. This newly discovered
fragility of magic numbers only becomes apparent far away from the
line of stability. Thus, exotic neutron-rich nuclei have opened a new
window into the elusive isovector sector of the nuclear energy density
functional (EDF). Moreover, some of these discoveries are providing
meaningful constraints on the behavior of neutron-rich matter, whose
equation of state (EOS) is essential for the understanding of complex
astrophysical objects such as core-collapse supernovae and neutron
stars. Such advances in terrestrial laboratories together with the
advent of powerful land- and spaced-based telescopes operating at 
a variety of wavelengths have created a unique and special synergy
between nuclear physics and astrophysics.

A ground-state observable that is highly sensitive to the EOS of
neutron-rich matter---particularly to the density dependence of the
symmetry energy---is the the neutron-skin thickness of ${}^{208}$Pb.
Indeed, despite the enormous difference in scales, an accurate
measurement of the neutron skin of ${}^{208}$Pb may provide vital
insights into the structure of neutron stars\,\cite{Horowitz:2000xj,
Horowitz:2001ya,Horowitz:2002mb,Carriere:2002bx,Steiner:2004fi,
Li:2005sr,Lattimer:2006xb,Lattimer:2012nd,Erler:2012qd,Fattoyev:2012rm,
Horowitz:2014bja}.  The Lead Radius Experiment (``PREX'') at the
Jefferson Laboratory has provided the first model-independent evidence
in favor of a neutron-rich skin in
${}^{208}$Pb\,\cite{Abrahamyan:2012gp, Horowitz:2012tj}. This
pioneering experiment---that will soon be upgraded to achieve the
originally proposed precision---measures a parity-violating asymmetry
in elastic electron scattering. This purely electroweak measurement is
a sensitive probe of neutron densities that is free from large and
uncontrolled strong-interaction uncertainties.  Moreover, PREX may
serve as an anchor to calibrate future hadronic measurements of 
neutron skins of exotic nuclei at rare isotope facilities.

Whereas the ground-state properties of exotic nuclei are of great
value in constraining the nuclear EDF, elucidating the full complexity
of the nuclear dynamics requires a comprehensive study of the response
of the nuclear ground state to a variety of probes. Indeed, nuclear
resonances offer a unique view of the nucleus that is often
inaccessible through other means\,\cite{Harakeh:2001}. In this new era of
radioactive beam facilities the study of novel modes of excitation in
exotic nuclei is a rapidly evolving area that holds great promise for
new discoveries\,\cite{Paar:2007bk}. Although interesting in their own
right, giant and pygmy resonances are also enormously valuable in
providing stringent constraints on the equation of state of asymmetric
matter\,\cite{Piekarewicz:2013bea}. In particular, the electric dipole
polarizability $\alpha_{{}_{\rm D}}$ has been shown to be highly
sensitive to the density dependence of the symmetry
energy\,\cite{Reinhard:2010wz,Piekarewicz:2010fa,Piekarewicz:2012pp,
Roca-Maza:2013mla}. This realization, in combination with a landmark
measurement of $\alpha_{{}_{\rm D}}$ in ${}^{208}$Pb at the Research
Center for Nuclear Physics\,\cite{Tamii:2011pv,Poltoratska:2012nf},
suggests that a comprehensive program of experimental measurements of
$\alpha_{{}_{\rm D}}$ on a variety of nuclei will place important
constraints on the isovector sector of the nuclear EDF.
 
Although primarily sensitive to the incompressibility coefficient of
symmetric nuclear matter, the isoscalar giant monopole
resonance---particularly in heavy nuclei with a significant neutron
excess---is also sensitive to the density dependence of the symmetry
energy because it probes the incompressibility of neutron-rich
matter\,\cite{Piekarewicz:2008nh}. Unfortunately, this sensitivity is
hindered by the relatively small neutron excess of the stable nuclei
measured up to date. Hence, measuring the isotopic dependence of 
the isoscalar giant monopole resonance (ISGMR) for both stable and
unstable nuclei is highly desirable. Thus, the recent report of a
measurement of the isoscalar monopole response of the unstable
neutron-rich ${}^{68}$Ni isotope by Vandebrouck and collaborators 
represents an important milestone\,\cite{Vandebrouck:2014}.

Pioneering experiments at GSI Helmholtzzentrum f\"ur
Schwerionenforschung have already measured the distribution of 
\emph{isovector dipole} strength in 
${}^{68}$Ni\,\cite{Wieland:2009,Rossi:2013xha}.  These experiments
have provided important insights into the emergence of low-energy
dipole strength in exotic nuclei and have also been used to constrain
critical parameters of the EOS, primarily the slope of the symmetry
energy at saturation
density\,\cite{Carbone:2010az,Piekarewicz:2010fa}.  In particular, the
identification of the electric dipole polarizabilty as a strong
isovector indicator by Reinhard and
Nazarewicz\,\cite{Reinhard:2010wz,Reinhard:2012vw} has provided an
observable whose precise experimental determination could reduce
theoretical uncertainties in the EOS. Given that the electric dipole
polarizability is proportional to the \emph{inverse} energy weighted
sum, the soft dipole mode (the so-called ``Pygmy'' dipole resonance)
plays a predominant role. Indeed, in the particular case of
${}^{68}$Ni, the soft dipole mode appears to exhaust as much as 25\%
of the total dipole polarizability\,\cite{Piekarewicz:2010fa}.
Thus, the soft dipole mode has generated considerable excitement as 
both a novel mode of excitation in exotic nuclei and as a possible
constraint on the EOS\,\cite{Tsoneva:2003gv,Piekarewicz:2006ip,
Tsoneva:2007fk,Klimkiewicz:2007zz,Carbone:2010az,
Papakonstantinou:2013gza}; for a recent comprehensive review on the
Pygmy Dipole Resonance see Ref.\,\cite{Savran:2013bha}.

However, it is essential to note that the origin of the distribution of 
low-energy strength in the isoscalar monopole and isovector dipole 
response is radically different. We now underline these 
differences in the context of the random-phase approximation (RPA)
framework that is used in most self-consistent calculations of the
strength distribution. Any RPA calculation starts by generating a
variety of ground-state properties that include single-particle
energies, the corresponding single-particle orbitals, and the
resulting mean-field potentials. With this information at hand, one
constructs the uncorrelated \emph{polarization propagator} that
consists of all particle-hole excitations with the spin and parity of
the excited state under consideration\,\cite{Dickhoff:2005}, i.e.,
$0^{+}$ for the monopole and $1^{-}$ for the dipole. In the particular
case of the monopole response, this involves $2\hbar\omega$
excitations connecting a particle to a hole with identical quantum
numbers (e.g., $2{\rm s}^{1/2}\!\rightarrow3{\rm s}^{1/2}$). Thus,
stable nuclei with strongly-bound orbitals display little (or no) 
uncorrelated low-energy monopole strength. This is in contrast
with dipole excitations that involve $1\hbar\omega$ excitations and 
thus particle and hole states that could be relatively close in energy 
(e.g., $2{\rm p}^{1/2}\!\rightarrow3{\rm s}^{1/2}$). Once the 
uncorrelated polarization propagator is obtained, the RPA response
emerges from the mixing of all relevant particle-hole excitation via 
a residual interaction that must be consistent with the one used to
generate the ground state. For isoscalar modes, the residual
particle-hole interaction is attractive and yields a collective
response that consists largely of a single fragment---the giant
monopole resonance---that exhausts most of the monopole
strength. Although the isovector residual interaction is repulsive,
thereby leading to a hardening and quenching of the response, for
stable nuclei the final outcome is similar, namely, a giant dipole
resonance that exhausts most of the classical energy weighted sum
rule. However, as the nucleus becomes neutron-rich, the isovector
interaction that played a relatively minor role in the ground-state
properties of nuclei with a small neutron excess, generates a
significantly repulsive contribution to the neutron mean-field
potential. This leads to weakly bound neutron orbitals that in some
cases may be close to the continuum.  Even so, in the case of the
dipole response there may be several discrete excitations in which
both the particle and the hole remain bound. In principle, these soft
excitations can be coherently mixed by the residual interaction and
ultimately generate a fairly well developed soft (pygmy) dipole 
resonance. 

However, the $2\hbar\omega$ character of the monopole excitations 
necessarily implies that all low-energy excitations must involve 
transitions into the continuum. Thus, a correct interpretation of 
the experimental results obtained by Vandebrouck and 
collaborators\,\cite{Vandebrouck:2014} requires a proper 
treatment of the continuum. In particular, the suggestion of a 
novel soft monopole mode---largely motivated by RPA predictions 
that use a discretized continuum---may be premature. Indeed,
it now appears that the prediction of low-energy monopole peaks 
that are well separated from the main giant resonance in the
neutron-rich Ni-isotopes\,\cite{Capelli:2009zz,Khan:2011ej} may 
be an artifact of the discretization\,\cite{Hamamoto:2014ala}. In a
recent analysis based on Skyrme-RPA calculations that do not involve
discretizing the continuum, Hamamoto and Sagawa conclude that 
``it is very unlikely to have some isoscalar monopole peaks with 
the width of the order of 1\,MeV below the excitation energy of 
20\,MeV in ${}^{68}$Ni''\,\cite{Hamamoto:2014ala}. In this contribution 
we report relativistic RPA calculations of the distribution of isoscalar
monopole strength with an exact treatment of the continuum for
${}^{56}$Ni, ${}^{68}$Ni, and ${}^{78}$Ni. Our results are in full
agreement with the conclusions by Hamamoto and Sagawa.

The paper has been organized as follows. In Sec.\,\ref{Formalism}
we briefly describe the relativistic RPA formalism used in this work,
with special emphasis on the treatment of the continuum. We then
proceed in Sec.\,\ref{Results} to display our predictions for the
distribution of isoscalar monopole strength in the isospin symmetric
${}^{56}$Ni nucleus as well as in the neutron-rich isotopes 
${}^{68}$Ni and ${}^{78}$Ni. Again, we pay special attention to
the role of the continuum in dictating the shape of the low-energy
strength. Finally, we end with a summary of our results in 
Sec.\,\ref{Conclusions}.

\section{Formalism}
\label{Formalism}

In this section we provide a brief review of both the formalism required
to compute the isoscalar monopole response and the physics that this
mode is sensitive to. In the case of the  response, special emphasis is 
placed on the importance of a proper treatment of the continuum.
Moreover, although it is well known that the nuclear ``breathing'' mode
probes the incompressibility coefficient of symmetric nuclear 
matter\,\cite{Blaizot:1980tw,Blaizot:1995zz,Harakeh:2001}, we stress 
that the monopole response of nuclei with a significant neutron excess 
is also sensitive to the poorly constrained density dependence of the 
symmetry energy.

\subsection{Isoscalar Monopole Response}
\label{IsMR}

The formalism underlying the calculation of the relativistic
mean-field ground state and the corresponding linear response 
has been reviewed extensively in earlier
publications\,\cite{Piekarewicz:2000nm,Piekarewicz:2001nm,
Piekarewicz:2013bea}. Although a full review of the formalism is no
longer necessary, we nevertheless highlight those points that are of
relevance to the present work, specifically the role of the continuum
on the distribution of low-energy isoscalar monopole strength.

In the relativistic mean-field (RMF) approach pioneered by Serot and
Walecka\,\cite{Walecka:1974qa,Serot:1984ey} the basic constituents of
the effective theory are protons and neutrons interacting via the
exchange of various mesons and the photon. In addition to these
conventional Yukawa terms, the model is supplemented by a variety of
nonlinear meson coupling terms that are critical to improve the
quality of the model\,\cite{Boguta:1977xi,Mueller:1996pm,
Horowitz:2000xj}.  In the RMF approximation, the nucleons satisfy a 
Dirac equation in the presence of strong scalar and vector potentials
that are generated by the meson fields. In turn, the mesons and the
photon satisfy classical Klein-Gordon equations with the relevant
nuclear densities acting as source term. Due to this close
interdependence, the equations of motion must be solved
self-consistently until convergence is
attained\,\cite{Todd:2003xs}. The self-consistent procedure culminates
with the determination of single-particle energies and Dirac
wave-functions, ground-state densities, and meson fields.

With such information at hand, one may proceed to compute the 
linear response of the mean-field ground state to a weak external
perturbation.  In the language of many-body theory, this requires 
the evaluation of the \emph{polarization
propagator}\,\cite{Fetter:1971,Dickhoff:2005}. The polarization
propagator---which is a function of both the energy and momentum
transfer to the nucleus---contains all dynamical information relevant
to the excitation spectrum of the system. Indeed, the polarization
propagator is an analytic function of the energy transfer except for
simple poles located at the excitation energies of the system and 
with the residues at the pole corresponding to the transition form
factor. The first step in the calculation of the isoscalar monopole 
response is the construction of the \emph{uncorrelated} 
(or mean-field) polarization propagator that is given by the
following expression:
\begin{align}
  \Pi_{ab}({\bf x},{\bf y};\omega) &= \sum_{0<n<F} 
   \!\overline{U}_{n}({\bf x})\gammai{0}\tau_{a}
    G_{F}\Big({\bf x},{\bf y};+\omega\!+\!E_{n}^{(+)}\Big) 
    \gammai{0}\tau_{b}\,U_{n}({\bf y})  \nonumber \\
    &+ \; \sum_{0<n<F} 
    \!\overline{U}_{n}({\bf y})\gammai{0}\tau_{b}
    G_{F}\Big({\bf y},{\bf x};-\omega\!+\!E_{n}^{(+)}\Big) 
    \gammai{0}\tau_{a}\,U_{n}({\bf x}) \,,
\label{Piab}
\end{align}
where $E_{n}^{(+)}$ and $U_{n}({\bf x})$ are the single-particle
energies and Dirac wave-functions obtained from the 
self-consistent determination of the mean-field ground state, 
$\gamma^{0}\!=\!{\rm diag}(1,1,-1,-1)$ is the zeroth component 
of the Dirac matrices, $\tau_{0}$ is the identity matrix in isospin 
space, and $\tau_{3}\!=\!{\rm diag}(1,-1)$ is the third isospin 
matrix. Note that the sum is restricted to positive-energy states 
below the Fermi level. Central to the calculation of the polarization 
propagator $\Pi_{ab}$ is the single-nucleon propagator $G_{F}$.  
Given that the ``Feynman propagator'' $G_{F}$ admits a spectral 
decomposition in terms of the mean-field solutions to the Dirac 
equation, its content is simple and illuminating. That is,
\begin{equation}
  G_{F}({\bf x},{\bf y};\omega) = \sum_{n}
   \left(
     \frac{U_{n}({\bf x})\overline{U}_{n}({\bf y})}
          {\omega - E_{n}^{(+)} + i\eta} + 
     \frac{V_{n}({\bf x})\overline{V}_{n}({\bf y})}
          {\omega + E_{n}^{(-)} - i\eta}  
   \right) \,,
 \label{GFeyn}
\end{equation}
where now $E_{n}^{(-)}$ and $V_{n}({\bf x})$ represent single-particle
energies and Dirac wave-functions associated with the negative-energy
part of the spectrum; recall that in the relativistic formalism the positive
energy part of the spectrum by itself is not complete. Note that the 
sum is now unrestricted, as it involves bound and continuum states of 
both positive and negative energy. Although the spectral decomposition
of the single-nucleon propagator is highly illuminating, its use in the 
calculation of the polarization propagator introduces certain artificial 
features---such as an energy cutoff and the discretization of the 
continuum---that may produce unreliable results. To avoid any 
reliance on artificial cutoffs and truncations, it is convenient to 
compute the nucleon propagator \emph{non-spectrally} by solving 
exactly for the relevant Green's function. That is,
\begin{equation}
  \Big(\omega\gammai{0}+i{\bm\gamma}\!\cdot\!{\bf\nabla}\!-\!M
  \!-\!\Sigma_{\rm MF}({\bf x})\Big)G_{F}({\bf x},{\bf y};\omega)=
  \delta({\bf x}-{\bf y}) \,,
 \label{GFeynEq}
\end{equation}
where ${\bm\gamma}\!=\!(\gammai{1},\gammai{2},\gammai{3})$
are Dirac gamma matrices and $\Sigma_{\rm MF}$ is the 
\emph{same exact} mean-field potential obtained from the 
self-consistent solution of the ground-state problem. Note that
it is only by ensuring that both the bound single-particle wave 
functions $U_{n}$ and the nucleon propagator $G_{F}$ ``move'' 
under the influence of the same mean-field potential that the 
conservation of the vector current can be 
ensured\,\cite{Piekarewicz:2001nm}. Moreover, the nonspectral 
approach has the enormous advantage that both the positive- and 
negative-energy continua are treated exactly. 

Having generated the mean-field polarization propagator, one 
proceeds to build coherence into the nuclear response by mixing 
all particle-hole excitations of the same spin and parity. Such a 
procedure is implemented by iterating the uncorrelated polarization 
propagator to all orders. The resulting RPA response often displays 
strong collective behavior that manifests itself in the appearance of 
one ``giant resonance" that exhausts most of the classical sum 
rule\,\cite{Harakeh:2001}. Besides its enormous impact on building 
the observed collectivity of certain nuclear modes, RPA correlations 
embody the correct self-consistent response of the mean-field ground 
state\,\cite{Thouless:1961,Dawson:1990wp,Ring:2004}. In particular,
in a seminal paper on vibrational states in nuclei, Thouless showed 
how spurious states---such as those associated with a uniform 
translation of the center-of-mass---separate out cleanly from the 
physical modes by shifting the spurious strength to zero excitation 
energy\,\cite{Thouless:1961}. In the context of the
relativistic formalism, Dawson and Furnstahl generalized Thouless' 
result by placing particular emphasis on the role of the negative-energy
states in the quest for consistency\,\cite{Dawson:1990wp}. 

\begin{figure}[ht]
 \vspace{-0.1cm}
 \begin{center}
  \includegraphics[width=0.40\linewidth,angle=0]{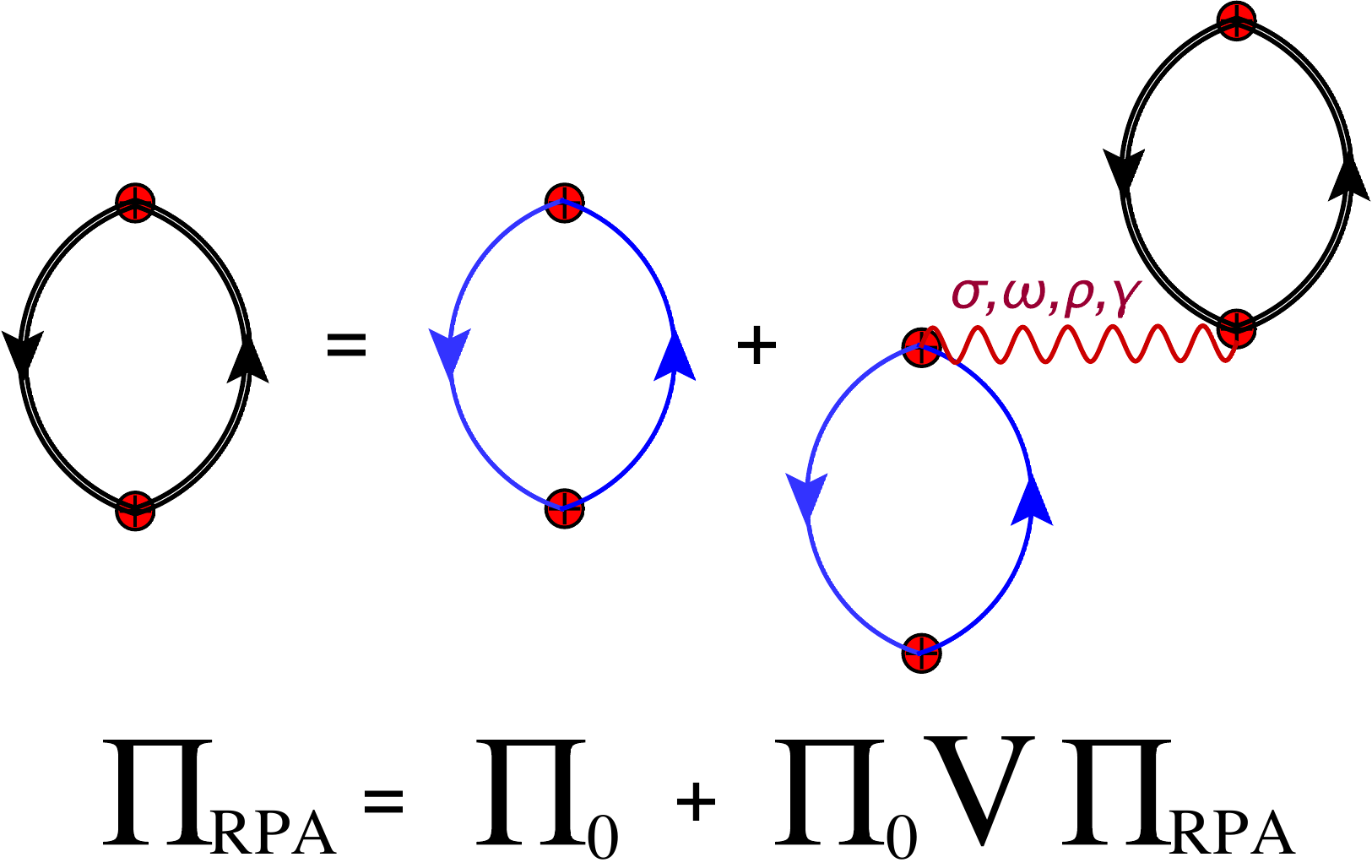}
  \caption{(Color online) Diagrammatic representation of the RPA (or Dyson's) 
  equations. The ring diagram with the thick black lines represents the fully 
  correlated RPA polarization while the one depicted with the thin blue lines is 
  the uncorrelated mean-field polarization. The residual interaction denoted with 
  the red wavy line must be identical to the one used to generate the mean-field 
  ground state.}
 \label{Fig1}
 \end{center} 
 \vspace{-0.25cm}
\end{figure}

By introducing the Fourier transform of the mean-field polarization 
propagator,
\begin{equation}
  \Pi_{ab}({\bf q},{\bf q}';\omega)\!=\!\!\int{d^{3}x}\,{d^{3}y}\, 
  e^{-i({\bf q} \cdot{\bf x}-{\bf q}'\cdot{\bf y})}\,
  \Pi_{ab}({\bf x},{\bf y};\omega) \,,
 \label{Piqq}
\end{equation}
one can obtain Dyson's equation for the RPA polarization whose solution 
encapsulates the collective response of the mean-field ground state. That 
is,
\begin{equation}
   \Pi_{ab}^{\rm RPA}({\bf q},{\bf q}';\omega)  =
     \Pi_{ab}({\bf q},{\bf q}';\omega) + 
   \int\!\frac{d^3k}{(2\pi)^{3}}\frac{d^3k'}{(2\pi)^{3}}
  \Pi_{ac}({\bf q},{\bf k};\omega)           
   V_{cd}({\bf k},{\bf k}';\omega)
  \Pi_{db}^{\rm RPA}({\bf k}',{\bf q}';\omega) \;,
 \label{PiabRPA} 
\end{equation}
where $V_{cd}({\bf k},{\bf k}';\omega)$ is the residual particle-hole interaction.
The diagrammatic representation of the RPA equations is displayed in 
Fig.\,\ref{Fig1}. It is worth repeating that the consistent linear response of the
system requires that both the mean-field potential $\Sigma_{\rm MF}$ and the 
residual particle hole interaction $V_{ab}$ be consistent with the interaction 
used to generate the mean-field ground state.

Finally, the distribution of isoscalar monopole strength may obtained by
taking the imaginary part of the polarization propagator projected into the 
correct ($J^{\pi}\!=\!0^{+},T\!=\!0$) channel. That is, 
\begin{equation}
  S(q,\omega;E0)\!=\!-
  \frac{1}{\pi} {\rm Im}\Big(\Pi_{00}^{\rm RPA}({\bf q},{\bf q};\omega)\Big) \,.
  \label{SRPA}
\end{equation}
In the long wavelength limit, the distribution of isoscalar monopole strength 
$R(\omega;E0)$ reduces to the following expression:
\begin{equation}
  R(\omega;E0) = 
  \lim_{q\rightarrow 0} \left(\frac{36}{q^{4}}\right) S(q,\omega;E0) \,. 
  \label{RGMR}
\end{equation}
In turn, moments of the distribution are defined as suitable energy weighted 
sums, namely,
\begin{equation}    
  m_{n}(E0) \equiv \int_{0}^{\infty}\!\omega^{n} R(\omega;E0)\, d\omega \,.
 \label{GMRMoments}
\end{equation}
Widely used in the literature are the energy weighted $m_{1}$, the energy unweighted 
$m_{0}$, and the inverse energy weighted $m_{-1}$ sums\,\cite{Harakeh:2001}. 

\subsection{Incompressibility of Neutron-Rich Matter}
\label{InNRM}

The saturation of infinite nuclear matter, namely, the existence of an equilibrium
density, is a hallmark of the complex and rich nuclear dynamics. Given that the
pressure vanishes at the equilibrium density, the small density fluctuations 
around saturation are described entirely by the incompressibility coefficient of 
symmetric nuclear matter $K_{0}$. Although $K_{0}$ can not be measured
directly in the laboratory, it can be tightly constrained from the strength distribution 
of the isoscalar monopole resonance. Indeed, using a sum-rule approach and
assuming that all the strength is concentrated in one collective peak, the
energy of the isoscalar giant monopole resonance (ISGMR) may be written
as follows\,\cite{Stringari:1982xx,Harakeh:2001}:
\begin{equation}
 E_{\rm GMR}=\sqrt{\frac{\hbar^{2} K_{A}}{M\langle r^{2}\rangle}} \,,
 \label{EGMR}
\end{equation} 
where $M$ is the nucleon mass, $\langle r^{2}\rangle$ the mean-square
nuclear radius, and $K_{A}$ the \emph{finite-nucleus} incompressibility
coefficient. It is important to note that although highly suggestive, modern 
theoretical approaches do not rely on the above expression to infer the
value of $K_{0}$. Rather, the same energy density functional is used to 
predict both $K_{0}$ and the distribution of isoscalar monopole strength.

Besides providing vital information on the equation of state of symmetric 
nuclear matter, the ISGMR of nuclei with a large neutron excess could 
shed light on the density dependence of the symmetry energy. Indeed, 
to the extent that the ISGMR probes the incompressibility of infinite
nuclear matter, the monopole response of nuclei with a significant 
neutron excess should be sensitive to the incompressibility 
coefficient of \emph{neutron-rich matter}\,\cite{Piekarewicz:2008nh}. To 
quantify the sensitivity of the symmetry energy to the ISGMR, we 
introduce the energy per particle of asymmetric matter at zero
temperature as follows:
\begin{equation}
  E/A(\rho,\alpha) - M \equiv {\cal E}(\rho,\alpha)
                          = {\cal E}_{\rm SNM}(\rho)
                          + \alpha^{2}{\cal S}(\rho)
                          + {\mathcal O}(\alpha^{4}) \,,
 \label{EOSa}
\end {equation}
where $\rho\!=\!\rho_{n}\!+\!\rho_{p}$ is the total (neutron plus proton) 
baryon density, ${\cal E}_{\rm SNM}$ is the energy per particle of 
symmetric nuclear matter, ${\cal S}$ is the symmetry energy, and 
$\alpha\!=\!(\rho_{n}\!-\!\rho_{p})/\rho$ represents the neutron-proton 
asymmetry. If one now expands the energy per particle around 
saturation density ($\rhoz$) one obtains
\begin{equation}
  {\cal E}(\rho,\alpha) = \Big(\epsz +
  \frac{1}{2}K_{0}x^{2}+\frac{1}{6}Q_{0}x^{3} +\ldots\Big) + 
  \alpha^{2}\Big(J + Lx + \frac{1}{2}K_{\rm sym}x^{2}
                 +\frac{1}{6}Q_{\rm sym}x^{3} +\ldots\Big) \,,
  \label{EOSb}
\end {equation}
where $x\!=\!(\rho\!-\!\rhoz)/3\rhoz$ describes the deviation of the density 
from its value at saturation. Here $\epsz$, $K_{0}$, and $Q_{0}$ denote
the binding energy per nucleon, curvature (i.e., incompressibility), and
skewness parameter of symmetric nuclear matter; $J$, $K_{\rm sym}$, 
and $Q_{\rm sym}$ represent the corresponding quantities for the
symmetry energy. Note, however, that unlike symmetric nuclear matter,
the symmetry pressure---or equivalently the slope of the symmetry 
energy $L$---does not vanish. This suggests that whereas symmetric 
nuclear matter ($\alpha\!\equiv\!0$) saturates at $\rhoz$, the presence
of $L$ modifies the saturation 
properties of asymmetric matter. Indeed, the following analytic 
expressions (correct to second order in $\alpha$) summarize the 
saturation properties of \emph{asymmetric nuclear 
matter}\,\cite{Piekarewicz:2008nh}:
\begin{subequations}
 \begin{align}
  & \rhoz(\alpha)=\rhoz + \rho_{\tau}\alpha^{2} 
    = \rhoz\left(1-3\frac{L}{K_{0}}\alpha^{2}\right) \,,
   \label{RhoTau} \\
  & \epsz(\alpha)=\epsz+\varepsilon_{\tau}\alpha^{2} 
    =\epsz+J\alpha^{2} \,,
   \label{ETau} \\
  & K_{0}(\alpha)=K_{0}+K_{\tau}\alpha^{2}  
  = K_{0}+\Big(K_{\rm sym}-6L-\frac{Q_{0}}{K_{0}}L\Big)\alpha^{2} \,.
 \label{KTau}
 \end{align}  
 \label{Taus}
\end{subequations}
Note that on very general grounds---both theoretical from the dynamics
of pure neutron matter\,\cite{Schwenk:2005ka,Hebeler:2009iv,
Gezerlis:2009iw,Vidana:2009is,Gandolfi:2009fj,Tews:2012fj,
Gandolfi:2013baa} and a variety of correlation
studies\,\cite{Brown:2000,
Furnstahl:2001un,Centelles:2008vu,RocaMaza:2011pm}, as well as
experimental from a measurement of the neutron-rich skin in
${}^{208}$Pb\,\cite{Abrahamyan:2012gp,Horowitz:2012tj}, the value of
the slope of the symmetry energy $L$ has been constrained to be positive.  
As a result,
Eq.\,(\ref{RhoTau}) indicates that neutron-rich matter saturates at
lower densities. Moreover, although slightly more uncertain, the
correction term to the incompressibility coefficient ($K_{\tau}$)
appears also to be negative, as it is dominated by the slope of the
symmetry energy\,\cite{Piekarewicz:2008nh,Piekarewicz:2013bea}; see
the large factor of 6 in front of $L$ in Eq.\,(\ref{KTau}). This
suggests that measurements of the isotopic dependence of the giant
monopole resonance---that should include unstable nuclei with a very
large neutron excess---could place significant constraints on the
density dependence of the symmetry energy.  Important first steps 
in this direction have been already taken by Garg and
collaborators\,\cite{Li:2007bp,Li:2010kfa,Patel:2012zd}.  In the
present paper we concentrate on the unstable neutron-rich isotopes 
${}^{68}$Ni and ${}^{78}$Ni with neutron-proton asymmetries of
$\alpha_{68}\!=\!0.18$ and $\alpha_{78}\!=\!0.28$, respectively.

\section{Results}
\label{Results}

Having provided the necessary framework to compute the distribution of 
isoscalar monopole strength and having discussed the physics that this 
mode is sensitive to, we now proceed to display our results using three 
relativistic mean-field models: (a) NL3\,\cite{Lalazissis:1996rd,Lalazissis:1999}, 
FSUGold\,\cite{Todd-Rutel:2005fa}, and IUFSU\,\cite{Fattoyev:2010mx}. 
Whereas both NL3 and FSUGold are accurately-calibrated interactions,
IUFSU involves a fine tuning of FSUGold in response to an interpretation 
of x-ray observations of neutron stars that suggest that FSUGold predicts 
stellar radii that are too large and a maximum stellar mass that is too 
small\,\cite{Steiner:2010fz}. Model parameters (i.e., coupling constants 
and meson masses) for these three sets of  interactions have 
been listed in Table I of Ref.\,\cite{Fattoyev:2010mx}.

Earlier attempts aimed at connecting the energies of the GMR to the
incompressibility coefficient $K_{0}$ relied heavily on dangerous
extrapolations from the properties of finite nuclei to those of
infinite nuclear
matter\,\cite{Blaizot:1980tw,Blaizot:1995zz}. However, as a result of
the much stricter standards imposed on the field today, predictions
for a variety of bulk properties of infinite nuclear matter as well as
the distribution of isoscalar monopole strength may now be provided
without any recourse to semi-empirical mass formulas. In
Table\,\ref{Table1} we list the predictions for a variety of bulk
properties of infinite nuclear matter at saturation density $\rhoz$ as
defined in Eq.\,(\ref{EOSb}).  In particular, $\epsz$, $K_{0}$, and
$Q_{0}$ represent the binding energy per nucleon, the
incompressibility coefficient, and skewness parameter of symmetric
nuclear matter, while $J$, $L$, and $K_{\rm sym}$ represent the
energy, slope, and curvature of the symmetry energy. From these
quantities, one can then obtain the asymmetric contribution to the
incompressibility coefficient $K_{\tau}$, as per Eq.\,(\ref{KTau}).
Finally, $K_{56}$, $K_{68}$, and $K_{78}$ represent the
incompressibility coefficient of neutron-rich matter having the same
neutron excess as ${}^{56}$Ni($\alpha\!=\!0$),
${}^{68}$Ni($\alpha\!=\!0.18$), and ${}^{78}$Ni($\alpha\!=\!0.28$),
respectively. Note that although the NL3 prediction for $K_{0}$ is
significantly larger than for the other two models, the differences
disappear almost entirely in the case of $K_{78}$; i.e., by the time
the neutron-proton asymmetry has grown up to $\alpha\!=\!0.28$. 
This is due to the very stiff symmetry energy of NL3 which, in turn,
provides a large correction to $K_{0}$, i.e., 
$K_{\tau}\!\approx\!-6L\!\approx\!-700$\,MeV. Thus, the  isotopic 
dependence of the ISGMR can help elucidate the density dependence 
of the symmetry energy---provided the isotopic chain includes exotic 
nuclei with very large neutron-proton asymmetries.

\begin{table}[h]
\begin{tabular}{|l||c|c|c|r|r|r|r||r|r|r|r|}
 \hline\rule{0pt}{2.25ex}   
 \!\!Model & $\rhoz$ & $\epsz$ 
           & $K_{0}$& $\hfill Q_{0}\hfill$& $\hfill J\hfill$& $\hfill L\hfill$& 
           $\hfill K_{\rm sym}\hfill$& $\hfill K_{\tau} \hfill$& $\hfill K_{56} \hfill$
           & $\hfill K_{68} \hfill$ & $\hfill K_{78} \hfill$\\
 \hline
 \hline\rule{0pt}{2.25ex} 
 \!\!NL3      &  0.148  & $-$16.24 & 271.5 & 209.5 & 37.29 & 118.2 &      100.9 &  $-$699.4 & 271.5 & 249.8 & 215.9\\ 
    FSU     &  0.148  & $-$16.30 & 230.0 & $-$522.7 & 32.59 & 60.5 & $-$51.3 & $-$276.9 & 230.0  & 221.4 & 208.0\\
 IUFSU  & 0.155  & $-$16.40 & 231.3 & $-$291.1 & 31.30 & 47.2  &     28.5 & $-$195.3 & 231.3  & 225.3 &215.8\\
\hline
\end{tabular}
\caption{Bulk parameters characterizing the behavior of infinite nuclear matter 
             at saturation density as defined in Eqs.\,(\ref{EOSb}) and\,(\ref{KTau}).
             Note that $K_{56}$, $K_{68}$, and $K_{78}$ 
             represent the incompressibility coefficient of asymmetric matter with
	      the same neutron excess as ${}^{56}$Ni, ${}^{68}$Ni, and ${}^{78}$Ni, 
	      respectively. All quantities are given in MeV except for $\rho_{{}_{0}}$ 
	      which is given in ${\rm fm}^{-3}$.}
\label{Table1}
\end{table}

Given that the self-consistent calculation of ground-state properties is
the necessary first step in the development of the RPA response, we
list in Table\,\ref{Table2} the predictions of all three models for the
binding energy per nucleon, root-mean-square charge and neutron
radii, and neutron-skin thickness of ${}^{56}$Ni, ${}^{68}$Ni, and
${}^{78}$Ni. To our knowledge, experimental values exist only for 
the binding energies\,\cite{Wang:2012}. 
Relative to experiment, the largest deviation in the binding energy 
is seen in the case of IUFSU: 1.7\% for ${}^{56}$Ni, 0.3\% for ${}^{68}$Ni,
and 1.5\% for ${}^{78}$Ni. For the charge radii, where experimental
measurements are not yet available, the spread among the predictions
amounts to less than half a percent for all three isotopes. However, as a
result of the large uncertainty in the value of the slope of the symmetry 
energy $L$, a marked discrepancy is observed in the predictions for the 
neutron radius and neutron-skin thickness of both neutron-rich nuclei. 
In the case of ${}^{78}$Ni, the difference between the stiffest (NL3) and
softest (IUFSU) models is about $0.13$\,fm. In particular, note that NL3 
predicts a very thick neutron skin of $R_{\rm skin}^{78}\!=\!0.42$\,fm. 

\begin{widetext}
\begin{center}
\begin{table}[h]
\begin{tabular}{|l||c|c|c|c||c|c|c|c||c|c|c|c|}
\hline\rule{0pt}{2.5ex} 
          &   \multicolumn{4}{ c|| }{${}^{56}$Ni}  
          &   \multicolumn{4}{ c|| }{${}^{68}$Ni}    
          &   \multicolumn{4}{  c| }{${}^{78}$Ni}   \\                         
 \hline
 \hline\rule{0pt}{2.5ex}  
\!\!Model  & $B/A$  & $R_{\rm ch}$ & $R_{n}$ & $R_{\rm skin}$ & 
                  $B/A$  & $R_{\rm ch}$ & $R_{n}$ & $R_{\rm skin}$  &
                  $B/A$  & $R_{\rm ch}$ & $R_{n}$ & $R_{\rm skin}$ \\                  
 \hline
 \hline\rule{0pt}{2.5ex}   
 \!\!NL3 & 8.608 & 3.701 & 3.578 & $-$0.049 
            & 8.688 & 3.855 & 4.045 & 0.261 
            & 8.239 & 3.934 & 4.281 & 0.416 \\     
 \hline\rule{0pt}{2.5ex}                
 \!\!FSU & 8.526 & 3.707 & 3.581 & $-$0.053 
            & 8.664 & 3.852 & 3.992 & 0.210 
            & 8.152 & 3.948 & 4.221 & 0.341 \\              
 \hline\rule{0pt}{2.5ex}                
 \!\!IUFSU & 8.501 & 3.680 & 3.553 & $-$0.053
            & 8.652 & 3.842 & 3.949 & 0.178 
            & 8.108 & 3.930 & 4.153 & 0.292 \\
\hline                  
\end{tabular}
\caption{Theoretical predictions for the binding energy per nucleon,
charge radius, neutron radius, and neutron-skin thickness of the 
three Nickel isotopes for NL3\,\cite{Lalazissis:1996rd,Lalazissis:1999}, 
FSUGold\,\cite{Todd-Rutel:2005fa}, and IUFSU\,\cite{Fattoyev:2010mx}. 
Experimental binding energies were extracted from the AME2012 
compilation\,\cite{Wang:2012}. Binding energies are given in MeV 
and radii in fm.}
\label{Table2}
\end{table}
\end{center}
\end{widetext}

\begin{figure}[ht]
\vspace{-0.05in}
\includegraphics[height=9.00cm,angle=0]{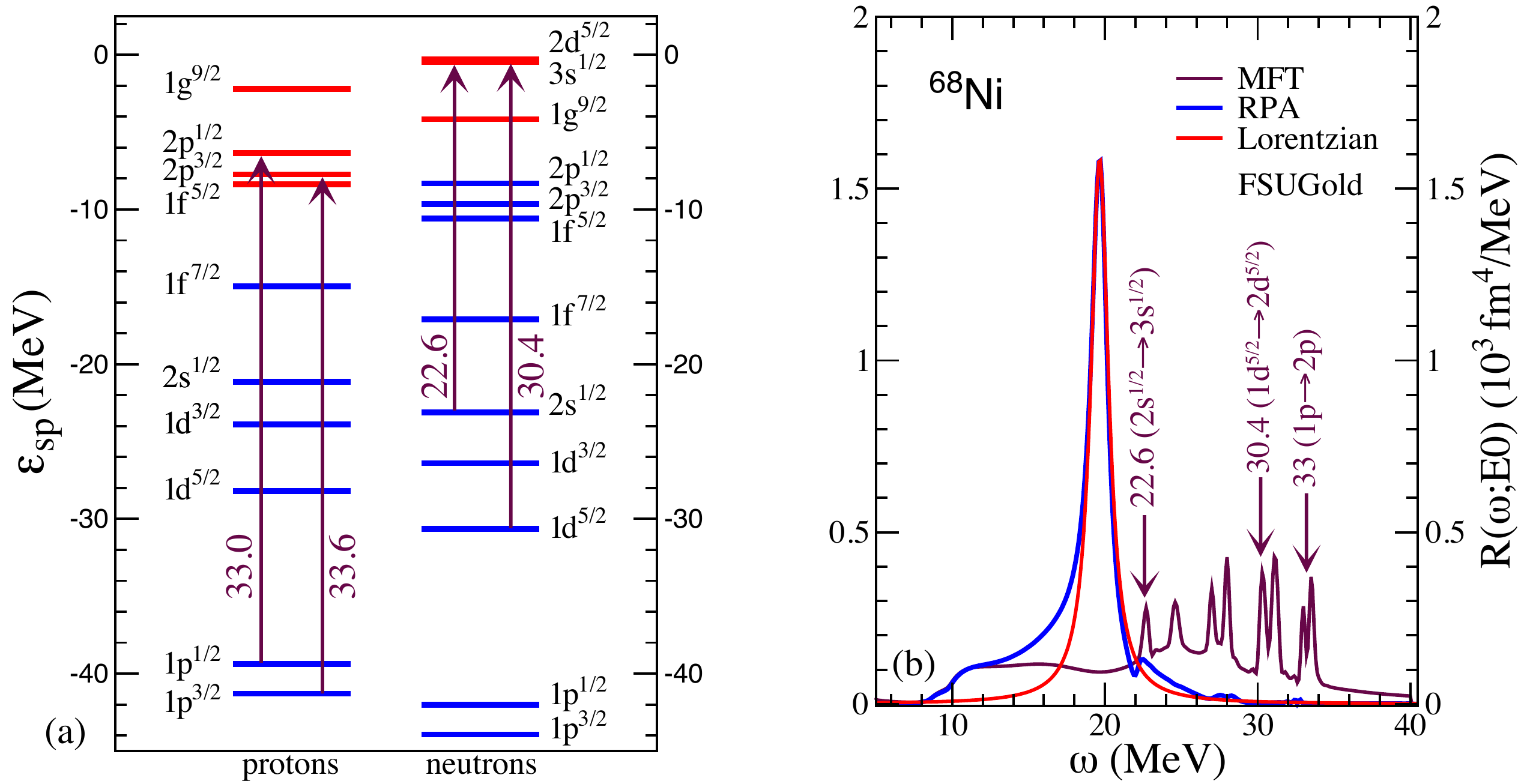}
\caption{(Color online) (a) Single-particle spectrum for ${}^{68}$Ni
as predicted by the relativistic FSUGold density functional. The
blue(red) lines denote occupied(empty) orbitals and the arrows are
used to indicate discrete excitations into bound states. (b)
Distribution of isoscalar monopole strength for ${}^{68}$Ni as
predicted by the relativistic FSUGold density functional.  Both
uncorrelated (MFT) and correlated (RPA) responses are displayed; 
the arrows indicate some of the expected mean-field transitions.
Also shown is a Lorentzian fit to the giant RPA peak to clearly
identify the excess strength at low energy.}
\label{Fig2}
\end{figure}

Having solved the self-consistent mean-field equations, which yield
single-particle energies and Dirac wave-functions as well as the
self-consistent scalar and vector mean-fields, one may now compute 
the distribution of isoscalar monopole strength in a relativistic
random-phase approximation [see Eq.\,(\ref{PiabRPA}) and
Fig.\ref{Fig1}]. We want to reiterate that the distribution of
monopole strength involves a \emph{non-spectral} solution of the
nucleon propagator that is free from any discretization of the continuum. 
However, in order to resolve discrete particle-hole
excitations, namely, excitations in which both the particle and the
hole are bound, we must supply the excitation energy $\omega$ 
with a small imaginary part of $\eta\!\equiv\!0.1$\,MeV. 

In Fig.\,\ref{Fig2}(a) we display FSUGold predictions for the
single-particle spectrum of ${}^{68}$Ni, with the arrows used to
indicate four prominent discrete excitations. As required, these four
discrete excitations are also clearly discernible in the distribution of
isoscalar monopole strength displayed in Fig.\,\ref{Fig2}(b). Moreover,
this uncorrelated (or MFT) response shows a significant amount of fairly
structureless strength from about 10 to 20 MeV followed by a series of
sharp peaks in the 20 to 35 MeV region. Note that most of the
``low-energy'' strength is generated by excitations from a bound Dirac
orbital into the continuum. As such, this component of the strength is
insensitive to the choice of $\eta$ (i.e., the small imaginary part of
$\omega$). In contrast, the sharp peaks at high-excitation energy
represent ``$2\hbar\omega$'' excitations that could not be resolved
without such a small imaginary part. Also note the presence of additional
``discrete'' peaks that are not identified in Fig.\,\ref{Fig2}(a). These 
extra peaks involve single-particle states at the edge of the continuum, 
such as the 3s${}^{1/2}$ and 2d${}^{5/2}$ proton orbitals.

The attractive isoscalar component of the residual interaction is
extremely efficient in mixing all individual particle-hole
excitations. This typically results in the development of a single
collective peak that exhausts most of the energy weighted sum
(EWS)\,\cite{Harakeh:2001}. The appearance of a giant monopole
resonance that carries most of the EWS is clearly discernable in
Fig.\,\ref{Fig2}(b) (blue solid line). This large collective mode is
sensitive to the incompressibility coefficient of infinite nuclear
and, at least for stable heavy nuclei, is accurately
described by a Lorentzian function. However, in the case of the 
exotic neutron-rich isotope ${}^{68}$Ni, a significant amount of
non-collective excess strength is observed at low energy. This fact 
is best illustrated by displaying (with a red solid line) a Lorentzian
fit to the large collective component. We attribute this difference
to the low-energy excitations into the continuum.  We note that 
the shape of the strength distribution at low energies is extremely 
sensitive to the treatment of the continuum, so one must exercise 
enormous care in drawing conclusions that rely on its 
discretization\,\cite{Hamamoto:2014ala}.

As we have just alluded, the nature of the low-energy strength is
associated with excitations from valence states into the continuum.
As the neutron excess increases, the isovector interaction---which in
the relativistic RMF model is dominated by vector exchange---becomes
repulsive for the neutrons and attractive for the protons. In the case
of ${}^{68}$Ni, this results in a closely-spaced triplet of orbitals
(1f\,${}^{5/2}$, 2p${}^{3/2}$, and 2p${}^{1/2}$) with a binding energy
of about 10\,MeV that hold the 12 extra neutrons relative to ${}^{56}$Ni 
[see Fig.\,\ref{Fig2}(a)]. This suggests that the emergence of low-energy
strength should closely track the neutron excess.  

\begin{figure}[ht]
\vspace{-0.05in}
\includegraphics[height=8.50cm,angle=0]{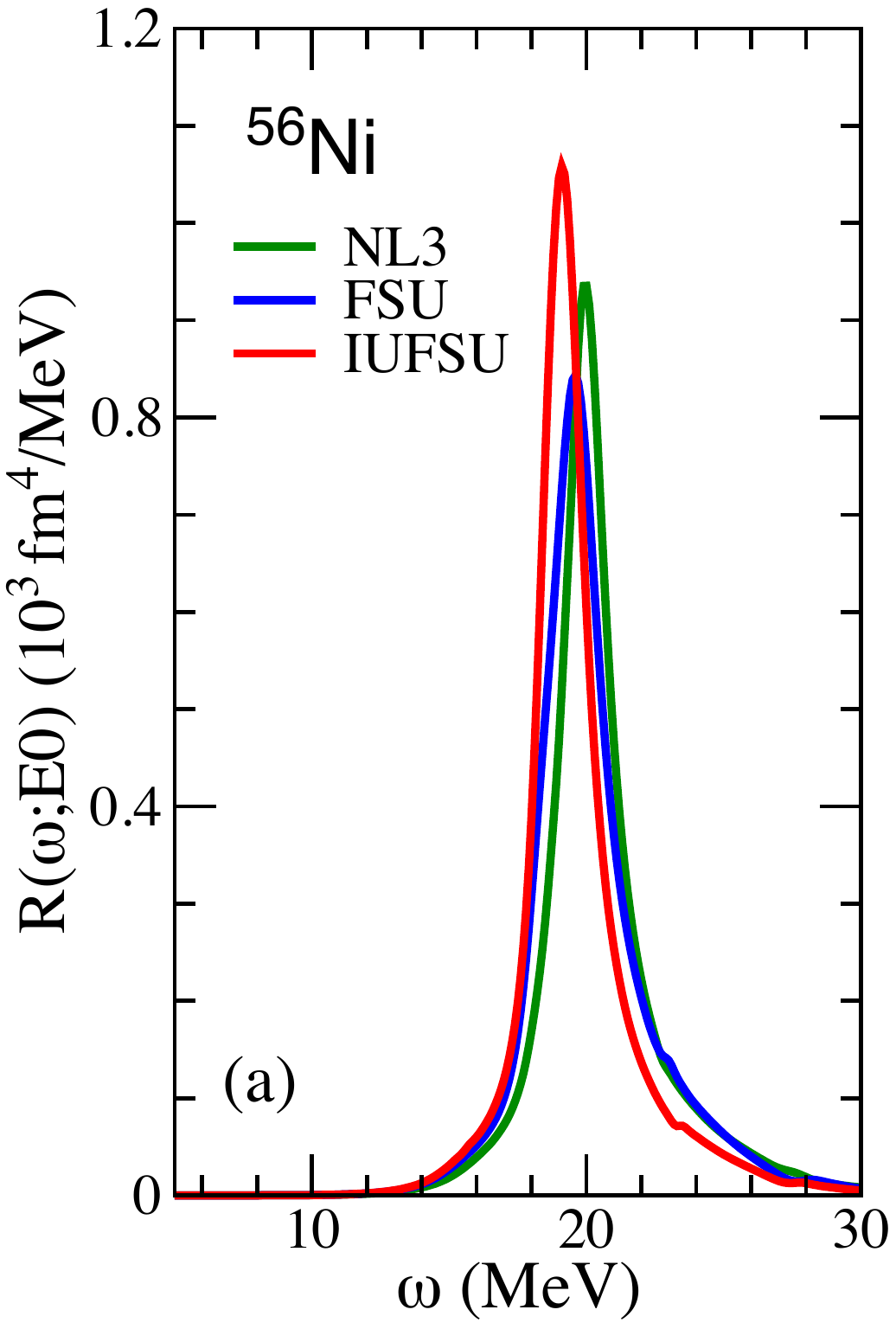}
 \hspace{0.2cm}
\includegraphics[height=8.50cm,angle=0]{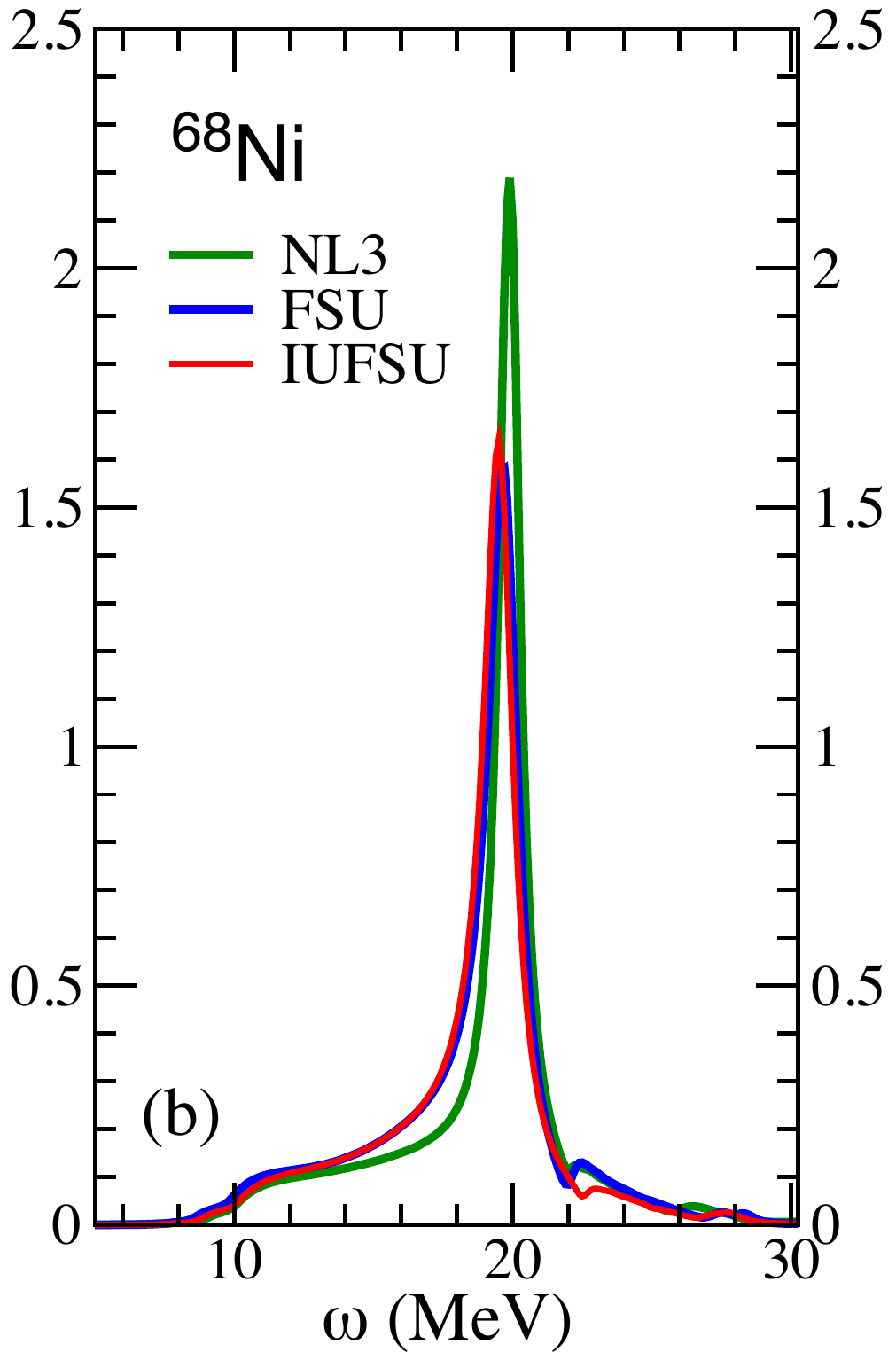}
 \hspace{0.2cm}
\includegraphics[height=8.50cm,angle=0]{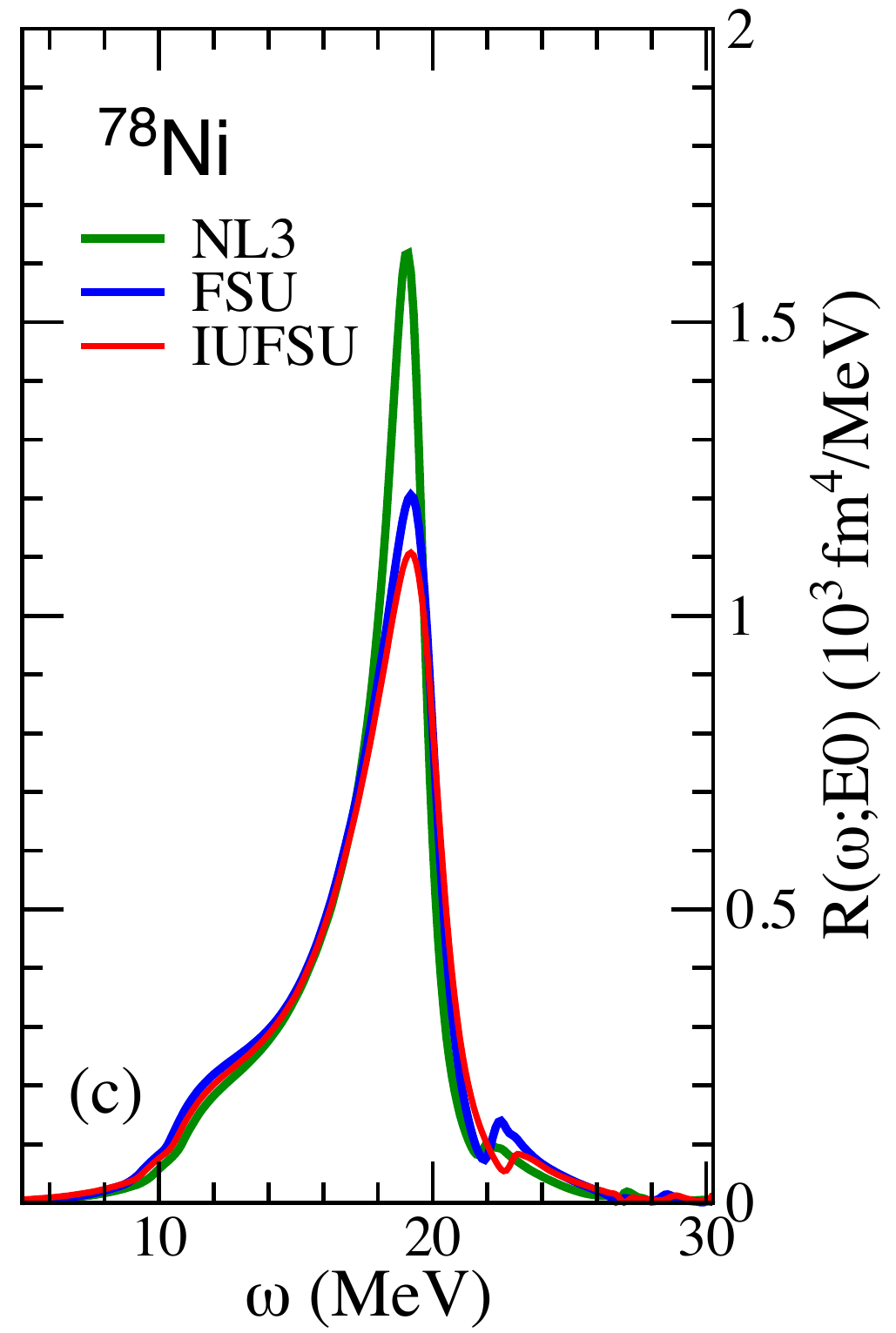}
\caption{(Color online) Distribution of isoscalar monopole strength 
for (a) ${}^{56}$Ni, (b) ${}^{68}$Ni, and (c) ${}^{78}$Ni as predicted by
relativistic RPA calculations using NL3\,\cite{Lalazissis:1996rd,Lalazissis:1999}, 
FSUGold\,\cite{Todd-Rutel:2005fa}, and IUFSU\,\cite{Fattoyev:2010mx}.} 
\label{Fig3}
\end{figure}

To test this assertion, we display in Fig.\,\ref{Fig3} the distribution of
isoscalar monopole strength for ${}^{56}$Ni, ${}^{68}$Ni, and
${}^{78}$Ni as predicted by relativistic RPA calculations. By
including three models with different bulk parameters, we can test the
sensitivity of the strength distribution to the density dependence of
the symmetry energy.  Also listed in Table\,\ref{Table3} are various
relevant moments of the distribution of strength that were obtained by
integrating from $\omega_{\rm min}\!=\!0.5$\,MeV to 
$\omega_{\rm max}\!=\!40$\,MeV; see Eq.\,(\ref{GMRMoments}).
The lack of low-energy strength in ${}^{56}$Ni is clearly
evident in Fig.\,\ref{Fig3}(a). Low-energy excitations into the
continuum that were driven by the neutrons occupying the
1f\,${}^{5/2}$, 2p${}^{3/2}$, and 2p${}^{1/2}$ orbitals in ${}^{68}$Ni
are absent in the case of ${}^{56}$Ni. Further, for a strength
distribution dominated by a single collective peak that is well
approximated by a Lorentzian shape, resonance energies satisfy the
following simple relations:
\begin{equation}
 \frac{m_{1}}{m_{0}} = \omega_{0} \quad{\rm and}\quad
 \sqrt{\frac{m_{1}}{m_{-1}}} =  \omega_{0}
 \sqrt{1+\frac{\Gamma_{0}^{2}}{4\omega_{0}^{2}}}
 \approx  \omega_{0} \,,
 \label{LorentzianMoments}
\end{equation}
where $\omega_{0}$ and $\Gamma_{0}$ represent the resonance energy 
and width, respectively. As expected, these relations are well satisfied
for the case of ${}^{56}$Ni. However, with the appearance of significant 
low energy strength in ${}^{68}$Ni---and even more so in
${}^{78}$Ni---significant distortions to the simple Lorentzian shape emerge. 

Given that the incompressibility coefficient for
symmetric nuclear is largest for NL3, its prediction for the GMR
energy of ${}^{56}$Ni is almost one MeV larger than for IUFSU (see
Table\,\ref{Table3}). Remarkably, however, the predictions of all 
three models for the distribution of strength in ${}^{78}$Ni are very 
similar to each other. This is consistent with the much faster softening 
of the incompressibility coefficient of neutron-rich matter for NL3 than 
in the case of either FSUGold or IUFSU; see the value of $K_{78}$ in
Table\,\ref{Table1}. Thus, studying the isotopic dependence of the
isoscalar monopole resonance for chains containing very exotic nuclei
may provide stringent constraints on the density dependence of the
symmetry energy\,\cite{Piekarewicz:2013bea}.

\begin{table}[h]
  \begin{tabular}{|c|c|c|c|c|c|c|c|c|}
    \hline\rule{0pt}{2.5ex} 
    \!\! Isotope & Model & $m_{-1}\,({\rm fm}^{4}/{\rm MeV})$ & $m_{0}\,({\rm fm}^{4})$ 
                   & $m_{1}\,({\rm fm}^{4}{\rm MeV})$ & $m_{3}\,({\rm fm}^{4}{\rm MeV}^{3})$ 
                   & $m_{1}/m_{0}\,({\rm MeV})$ & $\sqrt{m_{1}/m_{-1}}\,({\rm MeV})$ & 
                   $\sqrt{m_{3}/m_{1}}\,({\rm MeV})$ \\
   \hline
   \hline\rule{0pt}{2.5ex}   
    ${}^{56}$Ni &    NL3 & 0.138 & 2.810 & 58.031 & 26130.3 & 20.650 & 20.498 & 21.220 \\
                      &    FSU & 0.147 & 2.933 & 59.487 & 25850.1 & 20.283 & 20.130 & 20.846 \\
                      & IUFSU & 0.151 & 2.936 & 57.996 & 23763.5 & 19.752 & 19.622 & 20.242 \\
   \hline
   \hline\rule{0pt}{2.5ex}   
    ${}^{68}$Ni &    NL3 &  0.239 & 4.389 & 83.876 & 33659.6 & 19.112 & 18.751 & 20.032 \\
                      &    FSU &  0.254 & 4.524 & 84.045 & 32020.2 & 18.577 & 18.202 & 19.519 \\
                      & IUFSU &  0.248 & 4.433 & 82.080 & 30795.7 & 18.516 & 18.190 & 19.370 \\
   \hline
   \hline\rule{0pt}{2.5ex}   
    ${}^{78}$Ni &    NL3 &  0.358 & 6.037 & 105.999 & 35838.9 & 17.559	& 17.199 & 18.388 \\
                      &    FSU &  0.363 & 6.049 & 105.532 & 35687.6 & 17.446 & 17.042 & 18.389 \\
                      & IUFSU &  0.355 & 5.875 & 102.612 & 34849.4 & 17.466 & 17.002	& 18.429 \\
   \hline   
  \end{tabular}
 \caption{Various moments of the isoscalar monopole strength distribution (divided by a factor
 of $10^{3}$) and corresponding energies for the three Nickel isotopes considered in the text 
 as predicted by NL3\,\cite{Lalazissis:1996rd,Lalazissis:1999}, FSUGold\,\cite{Todd-Rutel:2005fa}, 
 and IUFSU\,\cite{Fattoyev:2010mx}. All moments were computed by integrating the distribution
 of strength from a minimum value of $\omega_{\rm min}\!=\!0.5$\,MeV to a 
 maximum value of $\omega_{\rm max}\!=\!40$\,MeV.} 
  \label{Table3}
 \end{table}

\section{Conclusions}
\label{Conclusions}
The unique and fascinating dynamics of exotic neutron-rich nuclei 
has lead to a paradigm shift in nuclear structure. Besides providing
insights into the limits of nuclear existence and the production of
heavy elements in the cosmos, the study of nuclei with large isospin
asymmetries opens a window into the poorly known nuclear isovector
interaction. In particular, large nuclei with a significant neutron
excess develop a neutron-rich skin that is highly sensitive to the
density dependence of the symmetry energy and, consequently, to the
nature of the isovector interaction.  Moreover, the electric dipole
polarizability and the development of low energy pygmy strength in
the isovector dipole response display strong sensitivity to the
isovector interaction. In the present contribution we have extended 
the study of the soft dipole mode and its sensitivity to the isovector 
interaction to the isoscalar monopole response of three magic 
(or semi-magic) Ni-isotopes---including the very neutron-rich
nuclei ${}^{68}$Ni and ${}^{78}$Ni.
 
The isotopic dependence of the isoscalar monopole resonance is of
great interest because the softening of the mode with increasing
neutron excess is highly sensitive to the density dependence of the
symmetry energy; see the expression for $K_{\tau}$ in Eq.\,(\ref{KTau}).
Although pioneering measurements of the isotopic dependence of
the ISGMR in both Tin and Cadmium have already been carried out, 
these measurements have been limited to the stables isotopes 
where the neutron excess, while significant, is not yet sufficiently 
large. Yet, we are confident that in the new era of rare isotope facilities 
these experimental studies will be extended much further. 
 
Central to our work was also the study of the emergence of low-energy
isoscalar monopole strength as a function of neutron excess. In the
case of the isovector dipole resonance, the appearance of low-energy
pygmy strength was seen to be strongly correlated to the development
of a neutron-rich skin in the Sn-isotopes. And although the nature of
the pygmy dipole resonance is still under debate, primarily whether it
is collective or not, the emergence of low-energy strength as a result
of a significant neutron excess is undeniable.

However, the nature of the isoscalar monopole strength at low energies
appears to be significantly more complex. To shed light on this
problem we have carried out relativistic RPA calculations of the
distribution of isoscalar monopole strength in ${}^{56}$Ni,
${}^{68}$Ni, and ${}^{78}$Ni.  In addition, we have used three RMF 
models with different assumptions on the isovector interaction to test 
the reliability of our conclusions. Finally and most importantly, our
RPA formalism is based on a non-spectral Green's function approach
where the continuum is treated on the same footing as the bound
states. This is in contrast to spectral calculations that must rely on
a discretization of the continuum. 

We conclude by summarizing our most important results. We find no low
energy monopole strength in the symmetric ${}^{56}$Ni isotope. Rather,
only one single collective giant resonance is identified with a centroid 
energy located at $m_{1}/m_{0}\!=\!20.65$\,MeV for NL3 (a model with 
an incompressibility coefficient of $K_{0}\!\approx\!270$\,MeV) and at
19.75\,MeV for IUFSU (with $K_{0}\!\approx\!230$\,MeV).
However, in contrast to the case of ${}^{56}$Ni, a significant amount
of low-energy strength is observed in both neutron-rich isotopes
${}^{68}$Ni and ${}^{78}$Ni---especially in the case of the latter. We
associate this fairly structureless strength to the
excitation of the extra 12 and 22 neutrons into the
continuum. In the absence of RPA correlations, the shape of
the mean-field response consists of featureless strength from about
10 to 20 MeV followed by discrete particle-hole excitations in the 20
to 35 MeV region. Once the attractive residual interaction is
incorporated, the coherence among all particle-hole excitations gives
rise to an RPA response that is significantly softened and
enhanced. This yields a smooth distribution of isoscalar monopole
strength that in addition to the giant resonance peak displays a
significant amount of low-energy strength. However, unlike some of the
results obtained using a discretized continuum, we found no pronounced
monopole states in the low-energy region that are well separated from
the giant monopole resonance\,\cite{Khan:2011ej}.  Instead,
our results support the recent \emph{continuum} calculations by
Hamamoto and Sagawa that report a ``broad shoulder'' of low-energy
monopole strength and question the appearance of isoscalar monopole
peaks below 20\,MeV\,\cite{Hamamoto:2014ala}. Given that a proper
treatment of the continuum is absolutely critical, the possible indication
of a soft monopole mode in ${}^{68}$Ni located at an energy of 
$12.9\pm1.9$\,MeV\,\cite{Vandebrouck:2014}, and largely motivated 
by the predictions of Ref.\,\cite{Khan:2011ej}, may need further 
verification and validation.

\begin{acknowledgments}
The author wishes to express his enormous gratitude to Professor Elias 
Khan for his support and hospitality during a recent visit to IPN-Orsay 
where some of the ideas presented in this contribution were first discussed. 
This material is based upon work supported by the U.S. Department of 
Energy Office of Science, Office of Nuclear Physics under Award Number 
DE-FD05-92ER40750.
\end{acknowledgments}

\bibliography{NiIsGMR.bbl}

\end{document}